\documentstyle[prl,aps]{revtex}
\begin{document}
\title{Two-level Hamiltonian of a superconducting
quantum point contact}

\author{ D. A. Ivanov$^{1,2}$ and M. V. Feigel'man$^1$}

\address{ $^1$ L.D.Landau Institute for Theoretical Physics, 117940
Moscow, Russia \\
$^2$ 12-127 M.I.T. Cambridge, MA, 02139 USA}
\date{August 4, 1998}

\maketitle

\begin{abstract}
In a superconducting quantum point contact, dynamics
of the superconducting phase is coupled
to the transitions between the subgap states. We
compute this coupling and derive the
two-level Hamiltonian of the contact.

\end{abstract}

\vspace{0.5cm}
One of the key features of superconducting quantum point contacts (SQPC)
\cite{KO,L,BH,SBW,AB,MR,AI,LG,IF}
is the existence of subgap states
(so-called Andreev states)
whose energies depend on the phase difference across the contact \cite{F,B}:
\begin{equation}
E(\alpha)=\pm \Delta\sqrt{1-t\sin^2{\alpha\over 2}},
\label{levels}
\end{equation}
where $\Delta$ is the superconducting gap, $t$ is the normal transparency
of the contact. Each transversal mode propagating through the
contact generates two such states (with opposite energies). Thus,
at energy scales less than $\Delta$, it is often convenient
to describe the contact as a set of two-level systems. Further we assume
for simplicity that we have only a single propagating
mode (and therefore only two subgap levels).

However, for describing dynamics of the contact at nonconstant $\alpha$,
the information on the energy spectrum (\ref{levels}) at each value of $\alpha$
is not sufficient. Mathematically speaking, we need a connection on
the bundle of Hilbert spaces over the circle of possible values of $\alpha$,
and, more specifically, the projection of this connection onto the two-level
subspace. Technically, it amounts to computing the ``dynamic'' matrix
element
\begin{equation}
I(\alpha)=\langle 0 | {\partial\over\partial\alpha} |1 \rangle,
\label{dynamic}
\end{equation}
where $| 0 \rangle$ and $| 1 \rangle$ are the two subgap states at a given
value of $\alpha$. This quantity defines coupling between the dynamics
of the superconducting phase and transitions in the two-dimensional subspace
of subgap states.

To illustrate this point, consider a single SQPC connected to a grain of
finite capacity $C$. This system has been studied previously in the
adiabatic approximation \cite{SZ,AL} and in the two-level approximation
\cite{IF,A}.
The two-level Hamiltonian for this system may be written as
\begin{equation}
H=H_0(\alpha) + {1\over 2C} \left( i{\partial\over\partial\alpha}
-N\right)^2,
\label{fullham}
\end{equation}
where $N$ is the dimensionless potential of the grain and $H_0(\alpha)$ is
a $2 \times 2$ matrix. At each $\alpha$, the eigenvalues of
$H_0(\alpha)$ must be given by (\ref{levels}).
However, this does not fix the whole
dependence on $\alpha$. In our earlier work \cite{IF} we suggested
\begin{equation}
H_0(\alpha)=\Delta \pmatrix{ \cos{\alpha\over 2}&\sqrt{r}\sin{\alpha\over2}
                           \cr
                         \sqrt{r}\sin{\alpha\over2}& -\cos{\alpha\over 2} }
\qquad (r=1-t).
\label{ham1}
\end{equation}
Another candidate for $H_0(\alpha)$ might be
\begin{equation}
H^\#_0(\alpha)=\pmatrix{E(\alpha) & 0 \cr 0 & -E(\alpha)}.
\label{ham2}
\end{equation}
Obviously, the latter choice of $H_0(\alpha)$ would lead to a physically
different behaviour of the system (\ref{fullham}), although $H^\#_0(\alpha)$
has the same eigenvalues as $H_0(\alpha)$. The choice (\ref{ham1})
of $H_0(\alpha)$
appears natural from the point of view of perturbation theory in
backscattering. Moreover, we claim that expression (\ref{ham1}) is exact
for any value of $r$ (within the model described below). We omitted the
derivation of (\ref{ham1}) in our paper \cite{IF} and now fill
this gap in the present
note.

We shall describe the one-channel SQPC by the one-dimensional Hamiltonian:
\begin{equation}
H_{full}=H_{SC}+H_{scatt}.
\label{formalham}
\end{equation}
\begin{eqnarray}
H_{SC}=\int_{-\infty}^{+\infty} dx \Big[ &&
i\Psi^\dagger_{L\beta} \partial_x \Psi_{L\beta}
-i\Psi^\dagger_{R\beta} \partial_x \Psi_{R\beta} + \nonumber\\
&& + \Delta(x)\big(\Psi^\dagger_{R\uparrow}\Psi^\dagger_{L\downarrow}
-\Psi^\dagger_{R\downarrow}\Psi^\dagger_{L\uparrow}\big) + \nonumber\\
&& + \Delta^*(x)\big(\Psi_{R\downarrow}\Psi_{L\uparrow}
-\Psi_{R\uparrow}\Psi_{L\downarrow}\big) \Big],
\label{scham}
\end{eqnarray}
where $\Psi^\dagger$ and $\Psi$ are electron operators ($L$ and $R$
subscripts denote left- and right-movers, $\beta=\uparrow,\downarrow$ is
the spin index), $\Delta(x)$ is the superconducting gap. We assume the
following coordinate dependence of the gap:
\begin{equation}
\Delta(x)=\cases{\Delta, & $x<0$ \cr \Delta e^{i\alpha}, & $x>0$ }
\end{equation}
in other words, the absolute value of the gap $\Delta$ is constant
across the contact, while the phase changes by $\alpha$ at $x=0$.

The scattering part of the Hamiltonian $H_{scatt}$ corresponds to elastic
scattering at $x=0$ and is also quadratic in electron operators.
We further disregard the nature of the scattering
and describe it by means of a scattering matrix.

The Hamiltonian is quadratic and may be diagonalized by operators linear
in $\Psi^\dagger$ and $\Psi$. We shall further compute the operators
$\gamma^\dagger_\uparrow$ and $\gamma^\dagger_\downarrow$ corresponding
to the subgap states. These operators satisfy the Bogolyubov-de-Gennes
equations \cite{deGennes}
\begin{equation}
\left[ \gamma^\dagger_\beta, H_{full}\right] = E\gamma^\dagger_\beta.
\label{bdg}
\end{equation}
\begin{eqnarray}
\gamma^\dagger_\uparrow&=&\int dx \Big[
u_L(x)\Psi_{L\downarrow}(x)+v_L(x)\Psi^\dagger_{R\uparrow}(x)
+u_R(x)\Psi^\dagger_{L\uparrow}(x)+v_R(x)\Psi_{R\downarrow}(x)
\Big], \nonumber\\
\gamma^\dagger_\downarrow&=&\int dx \Big[
u_L(x)\Psi_{L\uparrow}(x)-v_L(x)\Psi^\dagger_{R\downarrow}(x)
-u_R(x)\Psi^\dagger_{L\downarrow}(x)+v_R(x)\Psi_{R\uparrow}(x)
\Big].
\label{opers}
\end{eqnarray}
(Here we related $\gamma^\dagger_\uparrow$
and $\gamma^\dagger_\downarrow$ using
the spin-rotational invariance of the Hamiltonian).

Solving the equations (\ref{bdg}), we find that,
away from $x=0$, $u_\mu(x)$ and $v_\mu(x)$ have the following form:
\begin{eqnarray}
u_\mu(x)&=&\cases{u_\mu^+ e^{-\kappa x}, & $x>0$ \cr
                u_\mu^- e^{\kappa x}, & $x<0$}  \nonumber\\
v_\mu(x)&=&\cases{v_\mu^+ e^{-\kappa x}, & $x>0$ \cr
                v_\mu^- e^{\kappa x}, & $x<0$}
\label{uv}
\end{eqnarray}
where
\begin{equation}
\kappa=\sqrt{\Delta^2-E^2}.
\end{equation}
If we define
\begin{equation}
\phi=\arccos{E\over\Delta}=\arcsin{\kappa\over\Delta},
\end{equation}
the Bogolyubov-de-Gennes equations take the form:
\begin{eqnarray}
 e^{i\phi} u_R^- + v_R^- &=& 0, \nonumber\\
 e^{i\phi} u_L^- + v_L^- &=& 0, \nonumber\\
 e^{-i\phi} u_R^+ + e^{i\alpha} v_R^+ &=& 0, \nonumber\\
 e^{-i\phi} u_L^+ + e^{-i\alpha} v_L^+ &=& 0.
\label{bdg2}
\end{eqnarray}

The scattering matrix at $x=0$ matches $u_\mu(x)$ and $v_\mu(x)$
at $x=\pm 0$:
\begin{equation}
\pmatrix{u_L^- \cr v_R^+} =
\pmatrix{a & -b^* \cr b& a^*}
\pmatrix{u_L^+ \cr v_R^-}
\label{scatt1}
\end{equation}
\begin{equation}
\pmatrix{v_L^- \cr u_R^+} =
\pmatrix{a^* & -b^* \cr b & a}
\pmatrix{v_L^+ \cr u_R^-}
\label{scatt2}
\end{equation}
The former equation describes scattering of electrons, the latter one ---
scattering of holes. The amplitudes $a$ and $b$ must satisfy the unitarity
condition: $|a|^2 + |b|^2 =1$ ($|a|^2=t$, $|b|^2=r$).
The same scattering amplitudes in (\ref{scatt1}) and (\ref{scatt2})
only assume that the
scattering is spin-independent. We also neglect the momentum dependence
of the scattering amplitudes (the so-called ``instant scattering''
approximation). Finally, we remark
that the phase of $a$ has the same meaning as the superconducting phase
$\alpha$. Therefore, without loss of generality, we may assume that $a$ is
real: $a^*=a=\sqrt{t}$.

The condition that the homogeneous system of linear equations
(\ref{bdg2})--(\ref{scatt2})
has a solution, reduces to
\begin{equation}
\sin\phi = \sqrt{t} \sin{\frac{\alpha}{2}},
\end{equation}
which immediately gives (\ref{levels}) for the energy $E$ and
\begin{equation}
\kappa=\sqrt{t} \Delta \sin {\alpha\over 2}.
\end{equation}
(We assumed, without loss of generality, that $0\le \alpha\le\pi$ and
that $ E\ge 0$).

Solving the system (\ref{bdg2})--(\ref{scatt2}) enables us
to compute the following
commutator:
\begin{equation}
X(\alpha)=\left\{ \left[ {\partial H \over\partial\alpha},
\gamma^\dagger_\uparrow\right], \gamma^\dagger_\downarrow\right\}.
\end{equation}
We shall use this commutator to compute the ``dynamic'' matrix element
(\ref{dynamic}) as follows:
\begin{equation}
I(\alpha)=\langle 0 | {\partial\over\partial\alpha} |1 \rangle
= {1\over E_1-E_0}
\langle 0 | {\partial H \over\partial\alpha} |1 \rangle
={1\over 2E(\alpha)}
\langle 0 | {\partial H \over\partial\alpha}
\gamma^\dagger_\uparrow  \gamma^\dagger_\downarrow |0 \rangle
= {X(\alpha)\over 2E(\alpha)}.
\label{dynamic2}
\end{equation}
Since $H$ is quadratic in fermionic operators, $X(\alpha)$ is just a number.
A straightforward calculation (with normalized
$\gamma^\dagger_\uparrow$, $\gamma^\dagger_\downarrow$) gives
\begin{equation}
|X(\alpha)|=\sqrt{r} {\Delta^2\over 2E(\alpha)}.
\label{X}
\end{equation}

The phase of $X(\alpha)$ depends on the choice of phases of subgap
state operators $\gamma^\dagger_\uparrow$ and $\gamma^\dagger_\downarrow$
or, equivalently, on the relative phase of the
two states $|0\rangle$ and $|1\rangle$.
To fix this phase, we observe that our
system is invariant under the combined time reversal and
particle-hole symmetry. More specifically, this symmetry acts on operators
as follows:
\begin{eqnarray}
\Psi_{L\beta}& \mapsto & e^{i\xi}\Psi^\dagger_{R\beta}, \nonumber\\
\Psi_{R\beta}& \mapsto & -e^{-i\xi}\Psi^\dagger_{L\beta},
\end{eqnarray}
together with the complex conjugation of coefficients. The phase $\xi$
is adjusted depending on the phase of the backscattering amplitude $b$.
(This is a modified version of the well-known symmetry of
Bogolyubov-de-Gennes equations \cite{deGennes}).
If we choose the Andreev states to be self-conjugate,
then
\begin{equation}
\langle 0 | {\partial \over
 \partial\alpha} |0\rangle =
\langle 1 | {\partial \over \partial\alpha} |1\rangle = 0,
\qquad
I(\alpha)=
\langle 0 | {\partial\over\partial\alpha} |1\rangle {\rm ~is~real}.
\label{real}
\end{equation}
(In analogy to the ordinary quantum mechanics with a real
Hamiltonian: we may choose all eigenfunctions to
be real, then the matrix elements of real operators will also be real.)

With this choice of phases, from (\ref{dynamic2}),(\ref{X}),
\begin{equation}
I(\alpha)
=\sqrt{r} {\Delta^2\over 4E^2(\alpha)}.
\label{dynamic3}
\end{equation}

The operator of the charge on the grain, projected onto the two-dimensional
subbundle spanned by the states $|0\rangle $ and $|1\rangle$, takes
in the basis $\{|0\rangle,|1\rangle\}$ the form
\begin{equation}
Q=i\left[{\partial\over\partial\alpha} +
\pmatrix{\langle 0 |{\partial\over\partial\alpha} | 0\rangle &
        \langle 0 |{\partial\over\partial\alpha} | 1\rangle \cr
        \langle 1 |{\partial\over\partial\alpha} | 0\rangle &
        \langle 1 |{\partial\over\partial\alpha} | 1\rangle} \right]
=i\left[{\partial\over\partial\alpha} + I(\alpha)
\pmatrix{0 & 1 \cr -1 & 0} \right]
\end{equation}
The Hamiltonian (\ref{fullham}) in the basis $\{|0\rangle,|1\rangle\}$
takes the form:
\begin{equation}
H=\pmatrix{E(\alpha)&0\cr 0& -E(\alpha)}
+{1\over 2C}\left(i\left[{\partial\over\partial\alpha} + I(\alpha)
\pmatrix{0 & 1 \cr -1 & 0} \right] - N \right)^2.
\end{equation}

Now we want to perform the rotation to the ``fixed'' basis where
the charge operator is simply
\begin{equation}
Q=i{\partial\over\partial\alpha}.
\end{equation}
This results in the Hamiltonian (\ref{fullham}) with
\begin{equation}
H_0(\alpha)=U(\alpha)\pmatrix{E(\alpha)&0\cr 0&-E(\alpha)} U^{-1}(\alpha),
\label{rotation1}
\end{equation}
where $U(\alpha)$ is a rotation:
\begin{equation}
U(\alpha)={\rm P}\exp i\int I(\alpha) \pmatrix{0&1\cr -1& 0} d\alpha =
\pmatrix{\cos\varphi(\alpha) & -\sin\varphi(\alpha) \cr
\sin\varphi(\alpha) & \cos\varphi(\alpha)}.
\label{rotation2}
\end{equation}
The angle of rotation $\varphi(\alpha)$ is given by
\begin{equation}
\varphi(\alpha)=\int
\langle 0 | {\partial\over\partial\alpha} |1\rangle\, d\alpha =
\int {\sqrt{r}\over 4} {d\alpha\over 1-t \sin^2(\alpha/2)}
={1\over2}\arctan \left(\sqrt{r}\tan {\alpha\over2}\right)
\label{rotation3}
\end{equation}
which upon substituting in (\ref{rotation1})--(\ref{rotation2})
gives the result (\ref{ham1}).

To summarize, we have replaced the multi-body superconducting system
by the quantum-mechanical two-level Hamiltonian for the superconducting
phase across the contact.
This two-level approximation is appropriate whenever the system
stays away from the upper continuum of excitations. The Hamiltonian
(\ref{fullham}) loses its validity at points where the upper Andreev
state touches the upper continuum (at $\alpha=2\pi n$). Once a particle
reaches this point, it will pass to the vacant levels of the continuum
instead of following the localized subgap levels \cite{G}. This effect
is important for many non-equilibrium problems, for example, those
with constant voltage applied to the contact \cite{AB}. For
most equilibrium problems \cite{IF,A} and for some non-equilibrium
setups \cite{G,G2} the Hamiltonian defined by Eqs.~(\ref{fullham}),
(\ref{ham1}) may be used as the two-level approximation \cite{remark}.

This research was supported by the collaboration grant \#~7SUP~J048531
from the Swiss NSF, INTAS-RFBR grant \#~95-0302,
RFBR grant \#~98-02-19252, Program ``Statistical Physics" of the Russian
Ministry of Science, DGA grant \#~94-1189



\begin{references}
\bibitem{remark}
The Hamiltonian used in ref.~\cite{A} is incorrect away from the
transition region $\alpha\approx\pi$. It is appropriate
for the problem posed, but cannot be used for other problems involving
transitions away from $\alpha\approx\pi$. In such cases, the Hamiltonian
(\ref{fullham}),(\ref{ham1})
should be used instead.
\end{references}
\end{document}